\documentclass[conference]{IEEEtran}
\IEEEoverridecommandlockouts
\usepackage{cite}
\usepackage{amsmath,amssymb,amsfonts}
\usepackage{algorithmic}
\usepackage{graphicx}
\usepackage{textcomp}
\usepackage{xcolor}

\usepackage{amsmath}

\newcommand\blfootnote[1]{%
  \begingroup
  \renewcommand\thefootnote{}\footnote{#1}%
  \addtocounter{footnote}{-1}%
  \endgroup
}

\def\BibTeX{{\rm B\kern-.05em{\sc i\kern-.025em b}\kern-.08em
    T\kern-.1667em\lower.7ex\hbox{E}\kern-.125emX}}
\begin{document}

\title{Fairness in AI: challenges in bridging the gap between algorithms and law}

\author{\IEEEauthorblockN{Giorgos Giannopoulos}
\IEEEauthorblockA{\textit{
\textit{Athena Research Center}}\\
Maroussi, Greece \\
giann@athenarc.gr}
\and
\IEEEauthorblockN{Maria Psalla}
\IEEEauthorblockA{\textit{
\textit{Athena Research Center}}\\
Maroussi, Greece \\
mpsalla@yahoo.gr}
\and
\IEEEauthorblockN{Loukas Kavouras}
\IEEEauthorblockA{\textit{
\textit{Athena Research Center}}\\
Maroussi, Greece \\
kavouras@athenarc.gr}
\and
\IEEEauthorblockN{Dimitris Sacharidis}
\IEEEauthorblockA{\textit{
\textit{Universite Libre Brussels}}\\
Brussels, Belgium \\
dimitris.sacharidis@ulb.be}
\and
\IEEEauthorblockN{Jakub Marecek}
\IEEEauthorblockA{\textit{
\textit{Czech Technical University}}\\
Prague, Czech Republic \\
jakub.marecek@fel.cvut.cz}
\and
\IEEEauthorblockN{Germán M Matilla}
\IEEEauthorblockA{\textit{
\textit{Czech Technical University}}\\
Prague, Czech Republic \\
martige1@fel.cvut.cz}
\and
\IEEEauthorblockN{Ioannis Emiris}
\IEEEauthorblockA{\textit{
\textit{Athena Research Center}}\\
Maroussi, Greece \\
emiris@athenarc.gr}
}

\maketitle

\begin{abstract}
In this paper we examine algorithmic fairness from the perspective of law aiming to identify best practices and strategies for the specification and adoption of fairness definitions and algorithms in real-world systems and use cases. We start by providing a brief introduction of current anti-discrimination law in the European Union and the United States and discussing the concepts of bias and fairness from an legal and ethical viewpoint. We then proceed by presenting a set of algorithmic fairness definitions by example, aiming to communicate their objectives to non-technical audiences. Then, we introduce a set of core criteria that need to be taken into account when selecting a specific fairness definition for real-world use case applications. Finally, we enumerate a set of key considerations and best practices for the design and employment of fairness methods on real-world AI applications.


\blfootnote{Preprint. Accepted in Fairness in AI Workshop @ ICDE 2024.} 

\end{abstract}


\section{Introduction}
\label{sec:intro}

Fairness is a concept that transcends cultural, societal, and individual boundaries. It's a fundamental principle deeply ingrained in human consciousness, reflecting our innate sense of justice and equity. While the specific interpretation of fairness may vary across cultures and contexts, its essence remains constant: the equitable and just treatment of all individuals.
At its core, fairness represents the commitment to treat people impartially and without prejudice. It embodies the idea that everyone should have an equal opportunity to succeed, that rewards should be commensurate with efforts and contributions, and that discrimination and bias should be eliminated from all aspects of life.


Despite its universality, the concept of fairness is inherently complex and thus challenging to define precisely. It is traditionally defined as the quality of being fair, particularly emphasizing impartial treatment. However, this simplicity belies the intricate nature of fairness, as what is considered ``fair'' can be highly contingent on the specific context and the perspectives of different stakeholders.
It often involves balancing conflicting interests and values. What may seem fair to one person might not be seen the same way by another, depending on their perspective and circumstances. Yet, the pursuit of fairness remains an enduring and aspirational ideal, inspiring individuals and societies to continually strive for a more just and equitable world.


Fairness in the context of artificial intelligence (AI) represents a multifaceted and evolving objective. Its core purpose is to establish AI systems that consistently deliver unbiased, equitable decisions while avoiding the perpetuation or exacerbation of societal inequalities. 
Within the domain of AI, ``fairness'' serves as a linchpin term. It is prominently featured in nearly all frameworks and principles governing responsible and ethical AI development. These principles advocate for the meticulous design of AI systems that adhere to legal frameworks, respect human rights, uphold democratic values, and promote diversity. This approach seeks to ensure that AI actively contributes to the cultivation of a fair and just society, minimizing the risk of discriminatory outcomes.

Yet, achieving fairness in AI is far from straightforward. It necessitates a granular understanding of the multifaceted nature of fairness, acknowledging its dynamic and context-dependent characteristics. Furthermore, the pursuit of fairness in AI is a continual process, requiring ongoing vigilance and adaptation to evolving societal norms and ethical standards. Accordingly, it prompts the need for agile AI development, robust testing, and continuous refinement to ensure that AI technologies operate in harmony with the ever-evolving definitions and expectations of fairness in a rapidly changing world.


In our work, we identify the considerable gap between the legal and the algorithmic viewpoint of discrimination and fairness, as also done by prominent works on the intersection of the two disciplines \cite{Algorithmicxenidis21}. Consequently, we make an effort to bridge this gap, by incorporating literature material from both disciplines and presenting a discussion on their intersection, comprising, mainly, a set of criteria that need to be taken into account, when designing algorithmic fairness methods for real-world use cases, so as to adhere to respective legal requirements and perspectives.

\section{Discrimination in Law}
\label{sec:discr}

\subsection{Discrimination in the EU law}
\label{sub:eulawdiscr}

The aim of non-discrimination law is to allow all individuals an equal and fair prospect to access opportunities available in a society \cite{handbookeulaw}.
In the EU, the legal framework against discrimination has been shaped by both the laws of the Council of Europe and EU legislation. This framework is enshrined in both primary and secondary legal instruments.

\subsubsection{The law of the Council of Europe}
\label{sub:eulawcouncil}

The Council of Europe is an intergovernmental organization with the aim of, among other things, promoting human rights and social equality.
The European Convention for the Protection of Human Rights and Fundamental Freedoms, also known as the European Convention on Human Rights (ECHR) \cite{EuropeanCourtofHumanRights2010}, which was adopted by the member states of the Council of Europe in 1950, enshrines in Article 14 the prohibition of discrimination, stating specifically: ``\textit{The enjoyment of the rights and freedoms set forth in this Convention shall be secured without discrimination on any ground such as sex, race, colour, language, religion, political or other opinion, national or social origin, association with a national minority, property, birth or other status}''. With the Twelfth Protocol of the ECHR, signed in 2000, member states ``having regard to the fundamental principle according to which all persons are equal before the law and are entitled to the equal protection of the law'' and considering that they should ``\textit{take further steps to promote the equality of all persons through the collective enforcement of a general prohibition of discrimination}'', adopted a more recent provision, introducing a general prohibition of discrimination. This provision specifically states that ``\textit{the enjoyment of any right set forth by law shall be secured without discrimination on any ground such as sex, race, colour, language, religion, political or other opinion, national or social origin, association with a national minority, property, birth or other status}''.

The European Social Charter (ESC) (revised) of the Council of Europe\footnote{https://www.coe.int/en/web/european-social-charter} is another fundamental treaty of the Council of Europe for human rights. In Part V, Article E specifically refers to the non-discrimination clause during the enjoyment of the rights of the Charter. It states: ``\textit{The enjoyment of the rights set forth in this Charter shall be secured without discrimination on any ground such as race, colour, sex, language, religion, political or other opinion, national extraction or social origin, health, association with a national minority, birth or other status}''.

In addition to the above, there are many other Acts and specific Conventions that contain and have, as their fundamental principle, the prohibition of discrimination, which underpins all legislation of the Council of Europe.

\subsubsection{The law of the European Union}
\label{sub:eulaweu}

In the initial Treaties of the European Community, there is no mention of fundamental rights or their protection. The early regulations against discrimination were limited to a provision that prohibited gender discrimination in employment, and subsequently, other areas were regulated, such as pensions, pregnancy, and mandatory social security systems.
In 2000, the EU and its member states, recognizing that their policies could impact human rights, made the proclamation of the Charter of Fundamental Rights of the European Union\footnote{https://www.europarl.europa.eu/charter/pdf/text\_en.pdf}. Article 21 of the Charter addresses the prohibition of discrimination, stating: ``\textit{Any discrimination based on any ground such as sex, race, colour, ethnic or social origin, genetic features, language, religion or belief, political or any other opinion, membership of a national minority, property, birth, disability, age or sexual orientation shall be prohibited}''. Additionally, Article 20 guarantees equality before the law, Article 22 declares respect for cultural, religious, and linguistic diversity, and Article 23 ensures gender equality.

In the Treaty on European Union\footnote{https://eur-lex.europa.eu/legal-content/EN/TXT/?uri=CELEX\%3A12012M\%2FTXT}, which is based on the Maastricht Treaty [which was subsequently amended with the following treaties: Amsterdam Treaty (1997), Nice Treaty (2001), Lisbon Treaty (2007)], provisions regarding the prohibition of discrimination exist. Specifically, in Articles 2 and 3, it is mentioned that the Union ``\textit{is founded on the values of respect for human dignity, freedom, democracy, equality, the rule of law and respect for human rights, including the rights of persons belonging to minorities. These values are common to the Member States in a society in which pluralism, non-discrimination, tolerance, justice, solidarity and equality between women and men prevail}'' and that ``\textit{shall combat social exclusion and discrimination, and shall promote social justice and protection, equality between women and men, solidarity between generations and protection of the rights of the child}''.

Beyond the aforementioned EU constitutional texts, there are four Directives that address non-discrimination law and provide further specific regulations:

\begin{itemize}
    \item
        The Council Directive 2000/43/EC of 29 June 2000 implementing the principle of equal treatment between persons irrespective of racial or ethnic origin.
    \item
        The Council Directive 2000/78/EC of 27 November 2000 establishing a general framework for equal treatment in employment and occupation.
    \item
        The Council Directive 2004/113/EC of 13 December 2004 implementing the principle of equal treatment between men and women in the access to and supply of goods and services.
    \item
        The Directive 2006/54/EC of the European Parliament and of the Council of 5 July 2006 on the implementation of the principle of equal opportunities and equal treatment of men and women in matters of employment and occupation.

\end{itemize}

\subsubsection{Discrimination categories}
\label{sub:eulawcateg}

Non-discrimination law aims to allow all individuals an equal and fair prospect to access opportunities available in a society \cite{handbookeulaw}.
European non-discrimination law addresses two general types of discrimination: direct and indirect discrimination. Direct discrimination means that a person is treated less favorably based on a protected attribute (e.g. race and ethnicity, gender, religion and belief, age, disability, or sexual orientation) that they possess in matters of a protected sector (e.g. the workplace, provision of goods and services). Different groups receive different levels of protection. Direct discrimination is grounded in the Aristotelian postulate of treating ``like cases alike'' and treating ``different cases differently'' unless there is an objective reason not to do so. Equality achieved on these terms is also called ``formal equality,'' or the ``merit principle'' \cite{biaspreservationwachter2020bias}. 

Indirect discrimination occurs when ostensibly neutral provisions or practices, universally applied, disproportionately disadvantage individuals with specific protected characteristics, such as religion, disability, age or sexual orientation. This form of discrimination is evident when ostensibly fair rules unrelated to protected attributes affect a particular group disproportionately. The concept acknowledges that addressing existing inequalities may require differential treatment among groups, as justified by the principle of justified indirect discrimination. In such cases, pursuing a legitimate aim is acceptable if the mechanisms pass the ``proportionality test'', ensuring legal necessity and proportionate nature. Distinguishing itself from direct discrimination, this approach recognizes and addresses social challenges faced by protected groups, emphasizing the need for tailored strategies. Importantly, the specter of indirect discrimination is pronounced in Artificial Intelligence, machine learning, and automated decision-making, as these systems often rely on inferences that may inadvertently perpetuate biases and disparities among different demographic groups.

\subsection{Discrimination in the US law}
\label{sub:uslawdiscr}

In this subsection, we discuss the antidiscrimination law of the United States. We begin with a brief description of the legal system of the United States, and then delve into the specifics of antidiscrimination laws.

\subsubsection{The Legal Framework of the United States: Navigating American Jurisprudence}
\label{sub:uslawdiscrjuris}
 
The American legal system is a dynamic framework designed for justice and fairness. Operating under the U.S. Constitution, it consists of federal and state components, each with three branches—legislative, executive, and judicial—to ensure checks and balances. Led by the Supreme Court, the judiciary interprets laws and relies on common law tradition and precedent. 
 
\noindent \textbf{The Congress (legislature)}.
Congress, one of the three coequal branches of government, holds significant powers according to the Constitution, being the sole entity authorized to create new laws or modify existing ones. These laws, termed statutory laws, play a crucial role in the legal landscape. Given the evolving nature of society, these laws are often broadly formulated, allowing flexibility in interpretation. Congress delegates authority to federal agencies for implementation, with the courts providing interpretive functions and acting as a check on Congress's power.


\noindent \textbf{The Courts (judiciary)}.
In the United States, the court system plays a pivotal role within the common-law framework, allowing courts to establish legal precedents that guide future decisions. This precedent, formed through past cases, obligates judges to follow the reasoning applied in similar situations. Courts are not only responsible for interpreting statutory laws and the Constitution but also for shaping a body of case law, which holds comparable authority to other legal sources. 


\noindent \textbf{Federal Agencies (executive governance)}.
Federal agencies, specialized government bodies, are created through legislative or presidential action for purposes like resource management and national security. They play a crucial role in anti-discrimination efforts through rulemaking and law enforcement. While guidelines are non-binding, federal agencies are essential for statute effectiveness. Their political independence varies, with some operating within the executive and others having quasi-judicial enforcement powers. Collaborating with courts may face inefficiencies due to diverse legal sources and enforcement methods. 
 
\subsubsection{Anti-discrimination Laws}
\label{sub:uslawdiscrantilaws}
 
In this chapter, we reference anti-discrimination laws that may safeguard various rights and groups based on sex, age, race, disability, sexual orientation, gender, sex characteristics, religion and other characteristics.
Specifically:

\begin{enumerate}
    \item Title VII of the Civil Rights Act of 1964, which prohibits employment discrimination based on race, color, religion, national origin, or sex. It also forbids retaliation against those who report discrimination. Title VII addresses disparate treatment and disparate impact.
    \item The Equal Credit Opportunity Act (ECOA) is a federal law preventing discrimination in credit transactions. It applies to all credit extensions, including those to businesses, and prohibits discrimination. 
    \item Title VIII of the Civil Rights Act of 1968, the Fair Housing Act, prohibits discrimination in housing based on race, color, religion, sex, familial status, national origin, and disability. 
    \item Title VI of the Civil Rights Act of 1964 prohibits discrimination in any program or activity receiving federal financial assistance. It ensures that no person in the United States is excluded from participation, denied benefits, or subjected to discrimination under such programs. 
    \item The Pregnancy Discrimination Act of 1978 (``PDA''), an amendment to Title VII, prohibits discrimination based on pregnancy, childbirth, or related medical conditions. 
    \item The Equal Pay Act of 1963 (EPA), an amendment to the Fair Labor Standards Act, prohibits sex-based wage discrimination between men and women performing equal work in the same establishment. 
    \item The Age Discrimination in Employment Act of 1967 (ADEA) is a U.S. labor law prohibiting employment discrimination against individuals aged 40 or older, ensuring equal employment opportunities. 
    \item Title I of the Americans with Disabilities Act of 1990 (ADA) prohibits discrimination against qualified individuals with disabilities in the private sector and government. 
    \item Sections 102 and 103 of the Civil Rights Act of 1991 amend Title VII and the Americans with Disabilities Act to allow jury trials and awards for compensatory and punitive damages in cases of intentional discrimination.
    \item Sections 501 and 505 of the Rehabilitation Act of 1973 prohibit disability discrimination in the federal government. They require federal agencies to reasonably accommodate qualified employees or applicants with disabilities. 
    \item The Genetic Information Nondiscrimination Act of 2008 (GINA) is U.S. federal legislation protecting individuals from discrimination based on their genetic information in health insurance and employment. 
    \item The Pregnant Workers Fairness Act of 2022 (PWFA) mandates covered entities to provide reasonable accommodations to qualified workers with known limitations related to pregnancy, childbirth, or related medical conditions, unless it causes undue hardship. 
    \item The Immigration and Nationality Act of 1965 (``INA'') established a preference system prioritizing relatives, skilled professionals, and refugees. Abolishing quotas, the law aimed to attract those contributing significantly to the country's growth. 

\end{enumerate}

\subsubsection{Navigating the Complexity of Anti-Discrimination Measures: Sector-Specific Insights}
\label{sub:uslawsectorinsights}

As we observe, the management of discrimination and prejudice is anchored in a sector-specific approach—with distinct characteristics.

The initial observation underscores that selecting legislative safeguards for a specific and targeted right or group facing discrimination leads to enhanced and more comprehensive protection for that particular right or group. It becomes apparent that a broad legislative framework striving to shield all groups from discriminatory behaviors or uphold all rights may lack clarity and fail to incorporate the necessary safeguards for effective implementation of anti-discrimination measures.

Delving into anti-discrimination laws yields a second crucial insight: "\textit{even though laws are sector-specific, understanding discrimination is challenging when examining any one set of institutions in isolation.}" Discriminatory behavior is intricately molded by numerous factors each time, potentially impacting a broader array of rights or groups. Regarding the factors influencing the shaping of discriminatory behavior, it is evident that they encompass the historical evolution and treatment of the rights of specific groups or even the groups themselves.

Finally, when reflecting on the entire legal framework concerning non-discrimination, it is essential to acknowledge that "\textit{legal change is not the end of the road but, in some ways, the beginning.}" This statement carries a dual implication. On one hand, legal equality or protection against discrimination is insufficient in itself to fully redress societal and everyday equality. On the other hand, no legislative provision alone can comprehensively overcome all the enduring and historically ingrained discriminatory behaviors directed at specific groups and individual characteristics.

\subsubsection{Decoding Discrimination: Disparate Treatment vs. Disparate Impact}
\label{sub:uslawdiscrantidisparate}

Disparate treatment, a manifestation of intentional discrimination, denotes the intentional differential treatment of individuals based on specific characteristics, notably within the purview of employment law, where it stands as substantiated evidence of illegal discrimination, notably under Title VII of the United States Civil Rights Act. This discriminative paradigm is characterized by the bestowal of unequal treatment upon an employee predicated upon a protected characteristic, such as race or gender. To substantiate a disparate treatment claim, litigants must articulate intentional discrimination, thereby demonstrating that they experienced less favorable treatment stemming from motivation attributable to a protected characteristic. The establishment of causation assumes paramount significance in the evidentiary framework of disparate treatment, with two methodological avenues for substantiating it: either by manifesting that the protected characteristic served as a "motivating factor" in the genesis of the adverse decision or by elucidating that it constituted a "but-for cause."

The conceptual underpinnings of disparate treatment resonate with the societal understanding of discriminatory behavior, necessitating meticulous consideration of the potential alterations in actions if an individual's protected characteristic underwent modification, all the while maintaining the constancy of other case facts. In the United States legal context, disparate treatment finds its antithesis in disparate impact, where ostensibly neutral rules inflict prejudicial consequences upon specific protected groups. Title VII, a legislative bastion against discrimination, explicitly prohibits differential treatment of applicants or employees predicated upon their membership in a protected class. A disparate treatment transgression materializes when an individual is singled out and subjected to less favorable treatment due to an impermissible criterion, thereby raising probing inquiries into the discriminatory intent underpinning the actions of the employer.

Conversely, disparate impact represents unintentional discrimination that disparately affects a specified group. In contradistinction to disparate treatment claims, the onus of proving intent is obviated within disparate impact cases, where the analytical focus pivots toward the discernment of discriminatory effects resultant from ostensibly neutral practices.

The applicability of this theory extends to predictive algorithms, whose deployment may unwittingly result in the disproportionate exclusion of racial minorities from employment opportunities, irrespective of the absence of overtly intentional discriminatory practices.

Diverging from disparate treatment, disparate impact centers on practices that, though unintentional, engender a disproportionate impact on a protected class. This necessitates a stringent criterion of justification and avoidability, implemented through a methodological framework reliant on burden-shifting. Beyond its diagnostic function in uncovering latent intentional discrimination, disparate impact is grounded in a motivation for distributive justice. It seeks to mitigate unjustified inequalities in outcomes, thereby aligning with the paradigm of equality of opportunity. In essence, disparate impact mandates decision-makers to accord similar treatment to ostensibly dissimilar individuals, seeking redress for existing dissimilarities stemming from historical injustices and compensating for disadvantages incurred due to unjust causes.

In contrast, disparate treatment is emblematic of intentional employment discrimination, exemplified by discriminatory practices such as the exclusive testing of specific skills for certain minority applicants. Disparate impact materializes when ostensibly neutral policies, like uniform testing for all applicants, inadvertently result in a disproportionate adverse impact on a protected group, thus exemplifying unintentional discrimination wherein procedures ostensibly identical for all adversely affect individuals within a protected class.

\section{Prevalent algorithmic fairness definitions}
\label{biasdetect}


In this section, we briefly present, in a simplified, by example manner, a set of widely used algorithmic fairness definitions. Our goal is not an exhaustive survey on the field; there exist numerous works that nicely cover this task, indicatively \cite{10.1145/3194770.3194776, 10.1007/s00778-021-00697-y}.
We rather aim to intuitively communicate the main rationale of different fairness definitions to non-technical audiences, essentially facilitating the mapping to more legal/ethical concepts, such as \textit{equal treatment} versus \textit{equal outcome}. 
Since classification is mostly examined in the literature, we emphasize this ML task.
Specifically, we assume binary classification and denote the \emph{positive} and \emph{negative} classes as $+$, $-$, respectively. Let $Y$ be the \emph{actual class} of an instance, and let $R$ denote the classifier's \emph{prediction}; both $Y$, $R$ take values in $\{+,-\}$. We assume there exists a \emph{protected} attribute (e.g., gender) denoted as $A$; any other attribute is denoted as $S$. Finally, we denote as $Pr(R \: | \: E)$ the probability that the classifier outputs $R$ when $E$ holds.

\subsection{Demographic parity} 
Demographic Parity \cite{dwork2012fairness} states that the proportion of each segment of a protected class should receive the positive outcome at equal rates. Therefore, a classifier is fair if the following formula is satisfied:
\begin{equation}
Pr(R=+ \: | \: A=a )=Pr(R=+ \: | \: A=b ) \;\; \forall a,b \in A
\label{eq:parity}
\end{equation}

\emph{Example:} Suppose that we have 10 female and 20 male applicants for a job. If 10 males receive the outcome \emph{hire}, then we have a 50\% probability of males being hired. The model is considered fair if the probability of females receiving the outcome \emph{hire} is also 50\%, meaning that 5 females should be hired. If fewer than 5 females are hired, the model is biased against females, and if more than 5 females are hired, the model is biased against males. 

\subsection{Conditional statistical parity}
Conditional statistical parity \cite{costoffair17} also demands that subjects with both protected and unprotected characteristics should be equally likely to receive a positive classification prediction, but only when other legitimate factors are taken into account. This implies that demographic parity should hold, but only for specific subsets of the instances:
\begin{equation}
  \begin{aligned}
Pr(R=+ \: | \: S=s, A=a )=Pr(R=+ \: | \:S=s, A=b ) \\ 
\forall a,b \in A \; \forall s \in S
\end{aligned}
\label{eq:condparity}
\end{equation}

\emph{Example:} Suppose that we have 10 female and 20 male applicants for a job, and we know that 10 male applicants are young, while 6 female applicants are young. If 5 young males receive the outcome \emph{hire}, then we have a 50\% probability of young males being hired. The model is considered fair if the probability of young females to receive the outcome \emph{hire} is also 50\% meaning that 3 young females should be hired. If fewer than 3 young females are hired, the model is biased against females, and if more than 3 young females are hired, the model is biased against males.

\subsection{Equal opportunity} 
The definition of equal opportunity \cite{eqodds16} requires the positive outcome to be independent of the protected class A, conditional on Y being an actual positive. Therefore, equal opportunity is based on the predicted outcome and the actual outcome whereas demographic parity and conditional statistical parity are only based on the predicted outcome.
\begin{equation}
  \begin{aligned}
Pr(R=+ \: | \: Y=+, A=a )=Pr(R=+ \: | \:Y=+, A=b ) \;\; \\ 
\forall a,b \in A \;
\end{aligned}
\label{eq:eqopp}
\end{equation}

\emph{Example:} Suppose that we have 10 female and 20 male applicants for a job and we know that 10 male applicants are good matches for the job and 6 female applicants are good matches for the job. If 5 males that are good matches get the outcome \emph{hire}, then we have a 50\% probability of males being hired conditioned they are good matches. The model is fair if the probability of females that are good matches to get the outcome \emph{hire} is also 50\%, meaning that 3 females should be hired conditioned that they are good matches. If less than 3 females who are good matches are hired, then the model is biased against females and if more than 3 females who are good matches are hired, the model is biased against males.

\subsection{Equalized odds} 
This definition \cite{eqodds16} is more restrictive since it demands that individuals in protected and unprotected groups should have equal true positive rate and equal false positive rate, satisfying the formula: 
\begin{equation}
  \begin{aligned}
Pr(R=+ \: | \: Y=y, A=a )=Pr(R=+ \: | \:Y=y, A=b ) \;\;  \\ 
y \in \{+,- \} \;\; \forall a,b \in A \;
\end{aligned}
\label{eq:eqodd}
\end{equation}


\emph{Example:} Suppose that we have 6 female and 12 male applicants for a job and we know that 6 male applicants and 3 female applicants are good matches for the job, while other applicants are bad matches for the job. Furthermore, suppose that the model gives the outcome \emph{hire} to 9 applicants and the outcome \emph{no-hire} to the other 9 applicants. If the 6 males that are good matches get the outcome \emph{hire} and the other 6 males that are not good matches get the outcome \emph{no-hire}, then we have a 100\% probability of males being hired conditioned they are good matches and 100\% probability of males not being hired conditioned they are not good matches. The model is considered fair if the probability of females getting outcome \emph{hire} is 100\% conditioned that they are good matches and the probability of getting the outcome \emph{no-hire} is also 100\%. This means that the model should hire all the 3 females who are good matches and reject all the 3 females who are bad matches.

\subsection{Demographic Disparity}
This definition applies to each protected group independently and checks if the fraction of accepted outcomes is larger than the fraction of the rejected outcomes:
\begin{equation}
  \begin{aligned}
Pr(R=+ \: | A=a) \geq Pr(R=- \: | A=a) \;\; \forall a\in A \;
\end{aligned}
\label{eq:dd}
\end{equation}

\emph{Example:} Suppose that we have 10 female applicants. The model is fair towards females if it gives the outcome \emph{hire} to more females than it gives the outcome \emph{not-hire}. This means that if more than 5 females are rejected, then the model is unfair towards females. 

\subsection{Conditional Demographic Disparity} 
The Demographic Disparity definition \cite{cdc13} can be further refined to arrive at the Conditional Demographic disparity definition that conditions specific subgroups of the examined protected group.
\begin{equation}
  \begin{aligned}
Pr(R=+ \: |S=s, A=a) \geq Pr(R=- \: | S=s, A=a) \;\; \\
\forall a\in A \; \forall s \in S
\end{aligned}
\label{eq:cdd}
\end{equation}

\emph{Example:} Suppose that we have 100 female applicants that apply to 5 different jobs and suppose that the model gives the outcome \emph{hire} to 40 females and the outcome \emph{no-hire} to 60 females. The definition of demographic disparity will conclude that the model is unfair. However, it may be the case that all females are accepted in the first 4 jobs and all females are rejected in the fifth job. The conditional demographic disparity will conclude that the model is fair towards females conditioned they apply for one of the first 4 jobs and unfairly conditioned they apply for the fifth job.


\subsection{Counterfactual Fairness}
The last individual fairness definition \cite{CounterfactualFairness17} 
employs a different technique to capture the notion of the similarity of individuals. More specifically, the definition states that if the value of a sensitive attribute of an individual changes, then the outcome predicted by the model should remain the same.\\

\noindent \emph{Example:} A male individual received the outcome \emph{hire}. We change the gender of the male individual to female (adjusting other features to this change) and let the model predict again as if the individual was female from the beginning. If the result is again \emph{hire}, then the model is fair towards the individual, unfair otherwise.

\section{Criteria for the selection of fairness methods}
\label{sub:criteria}

In this section, we briefly discuss a series of criteria that we have drafted for assessing the suitability of fairness policies and bias detection methods. These criteria have been formulated considering 
substantial risks, gaps, and issues identified in the literature at the intersection of ethics, law, and algorithmic fairness in decision-making, such as indirect/proxy discrimination, intersectional discrimination, and other related concerns.


\subsection{Equal treatment vs equal outcome}

Title VII of the Civil Rights Act of 1964 (USA)\footnote{https://www.eeoc.gov/statutes/title-vii-civil-rights-act-1964}
\cite{raceawareusakim2022} prohibits discrimination in employment based on the sensitive attributes of race, color, religion, sex, and national origin. 
In this setting, a distinction is made with respect to the notion of equality a policy, practice or method aims to achieve \cite{equaloppoutrushefsky1996public}. In particular:
\begin{itemize}
    \item \textbf{Equal treatment} prescribes that all individuals are given the same chances to achieve a favorable outcome. In our example, this means that a recommendation system should assign an ``accept'' label to candidates based on objective criteria and labeled training data, that do not take into account sex information in its decision.

    \item \textbf{Equal outcome} prescribes that all protected (sub)groups equally/proportionally obtain the favorable outcome. In our example, the ratio of males to females that obtain the ``accept'' label should be proportionate to the respective candidate's ratio, even if this conflicts with the rating produced by the recommender.

    Equal outcome is a notion that is based on the recognition of \textbf{structural} inequalities and \textbf{historical} bias in procedures and datasets and is achieved via instruments such as \textbf{affirmative action}\footnote{https://plato.stanford.edu/entries/affirmative-action/} (or \textbf{positive action}, \textbf{positive discrimination}), which aim at alleviating/eliminating such inequalities against sensitive subpopulations. In our case, affirmative action or a company's policy would require a minimum quota in female ``acceptances'' for every job.
\end{itemize}

We note that definitions $A$, $B$, $E$ and $F$, align with equal outcome, while $C$ and $D$ with equal treatment. Definition $G$ comprises a middle ground between the two concepts and, if appropriately applied, could achieve substantive equality, meaning equal treatment but taking into account and accounting for historical biases. 

When designing and developing fairness auditing and correction tools for a specific use case, it is crucial that the aforementioned dimensions are very clear, so that respectively appropriate fairness definitions are selected and properly configured. In particular, questions such as: ``\textit{is structural bias recognized in the specific use case? If so, are there directives, in the form of positive actions, that impose specific quota? Are there specific sensitive attributes that are highly relevant/informative features for an AI decision making system on the case and need to be taken into account and, vice versa, other ones than need to be ignored?}'' The latter question, although relevant to disparate impact, also highly relate to the second criterion, as discussed next.


\subsection{Handling of proxy variables and correlations - indirect discrimination}
A substantial issue when it comes to auditing discrimination, algorithmic or not, lies in cases of \textbf{proxy discrimination}. This is the case where bias in the data, or a trained ML model, is expressed not directly via sensitive attributes, but indirectly, via proxy variables, that are to some extent correlated with the respective sensitive attribute \cite{Barocas2016BigDD, DiscriminationAwareDataMining08}. Examples of indirect/proxy discrimination can be found in \textit{height} and \textit{maternity leave} attributes serving as proxies for the \textit{sex} sensitive attribute, and \textit{residence} or \textit{location} attributes serving as proxies for the \textit{race} sensitive attribute. Proxy discrimination is highly relevant to disparate impact, happening in cases where even though there exists no explicit discrimination on grounds of any sensitive attributes, and the decisions are taken via facially neutral practices, nevertheless specific protected groups end up being (disproportionally) disadvantaged.

Due to the commonly encountered misunderstanding that, upon sensitive attributes are excluded from an AI model's training, fairness is ensured (also called \textit{fairness by unawareness} \cite{CounterfactualFairness17}), bias can be perpetuated via proxy discrimination. That is, even if sensitive attributes are removed, the bias of the training data can still be transferred into the trained model: the ML algorithm will try to learn patterns that relate the labels of the data, that express this bias, with remaining features that correlate with the removed sensitive attribute. Take for example a training dataset on hiring, that is significantly biased against female individuals, i.e., in a binary classification setting (\textit{hire} or \textit{no-hire}), for very similar jobs, almost exclusively male individuals are hired and female ones are rarely hired. Suppose that the model owner removes the sex attribute from each individual and uses the training dataset to train a binary classifier for recommending whether to hire or not a new applicant, whose sex is, of course, also unknown. While the sex attribute is absent from each individual, there most probably exist other attributes that are correlated with it, such as \textit{university name} or \textit{years of experience after graduation}. Specific values for these attributes will also be correlated with the biased labels. For example, individuals from certain universities, which are traditionally attended by proportionally more female students, will mostly be assigned the \textit{no-hire} label in the training dataset. The ML algorithm, during its training, will learn such patterns, transferring this implicit bias into the trained model. Thus, for a new individual candidate, even if the sex is unknown, if they have attended the specific universities, they are most probably going to be assigned the biased, \textit{no-hire} label.

The above also serves as an example of a related issue of implicit bias, termed as \textit{discrimination by association} \cite{Wachter2019AffinityPA, Algorithmicxenidis21}. This issue appears when individuals are mistakenly categorized as part of a protected group, which faces discrimination, and consequently experience the same type of discrimination. In our example, the training data, and the derived ML model are biased towards female individuals and, by correlation, also towards individuals that have attended specific universities, even if they are males.\\



\subsection{Handling of intersectional/subgroup fairness}
Another important issue arises when considering subpopulations defined by more than one attribute, with at least one of them being a sensitive one. In such cases, \textit{subgroup fairness} \cite{preventingfairnessgerry17} or \textit{intersectional discrimination}\footnote{https://www.coe.int/en/web/gender-matters/intersectionality-and-multiple-discrimination} or multi-dimensional discrimination \cite{multidimntoutsi23} is audited. Consider for example a scenario where an AI system decides whether a person is promoted based on a specific feature set and we want to audit the fairness of the system concerning two sensitive attributes: \textit{gender}, with values $\{$\textit{male}, \textit{female}$\}$ and \textit{race}, with values $\{$\textit{Caucasian}, \textit{non-Caucasian}$\}$. In our scenario, the unprotected groups are defined by \textit{female} and \textit{non-Caucasian} respectively. It might be the case that auditing fairness individually on the two sensitive attributes finds the promotion decisions of the system fair, however, by further examining the result, one might identify that non-Caucasian males and Caucasian females are disproportionally unfavored compared to the other two subgroups, i.e., 
Caucasian males and non-Caucasian females. 

Subgroup bias presents quite a few challenges. First of all, it is often the case that bias is magnified for specific subgroups; this is not only due to preexisting discrimination towards individuals belonging to these groups but also due to their under-representation in datasets and models. That is, since these groups comprise only a very small subset of the general population, they are often not properly represented in sampling/training data-gathering processes. Another issue related to this data sparsity is the uncertainty in evaluating bias for these subgroups when auditing a dataset or algorithm: since very few instances representing a specific subgroup might be found in an audited dataset, the significance of the findings can be questionable. Finally, computational issues arise when trying to drill down to more granular subgroups, since complexity increases exponentially \cite{preventingfairnessgerry17}.


\subsection{Handling of feedback loops}

Feedback loops comprise self-repeating processes that can potentially reinforce and perpetuate preexisting bias \cite{Kim2017DataDrivenDA}. Consider our running example of a hiring recommendation system. If such a system is initially trained on a biased dataset, then its recommendations will probably reproduce the bias (if no fairness-correcting action is taken). Then, these new recommendations can be used as additional training data, that also carry bias. Further, applying the system in real-world domains and continuously rejecting female candidates in favor of male ones, might discourage individuals from the formerly protected groups from applying for specific job positions. It is well recognized in the literature that pattern recognition and learning mechanisms applied by many AI systems can facilitate the creation of feedback loops  \cite{Kim2017DataDrivenDA, Barocas2016BigDD, Algorithmicxenidis21}.



\subsection{Robustness to manipulation}

Discrimination can often be intentional, meaning that the system/data/application owner is aware of preexisting bias and does not apply any bias-correction actions or even tries to hide it, or explicitly introduces bias. Masking bias can be achieved through various ways, including gerrymandering \cite{youshouldnttrustDimanov2020YouST}, as discussed above, as well as by manipulating the output of fairness auditing/explainability methods to render the audited model seemingly fair, while it is not. The work of \cite{youshouldnttrustDimanov2020YouST} prominently demonstrates how a classifier can be retrained in an adversarial way, to maintain the same level of accuracy, and at the same time suppress the explicit contribution of sensitive attributes, so that a large set of explainability methods are tricked into falsely deciding that its outputs are fair.

\subsection{Sampling requirements}



AI systems typically require huge training datasets, where bias detection needs to be performed,
for instance, in terms of underrepresentation of some of the subgroups of the general population. 
There, one can compare the distribution of a protected attribute in the general population against the distribution of the protected attribute in the training data. 
Then, bias detection involves calculating distances between two probability distributions derived from our data. 
There exist various distances to employ, including \emph{Hellinger, Total Variation (TV), Wasserstein (OT), Maximum Mean Discrepancy (MMD)}, etc. These are expected to be calculated with an accuracy increasing in the number of samples considered in the training data (or subsample thereof used for the bias estimation). 
The relationship between the number of samples, and the error in estimating the bias is known as the sample complexity of bias detection.

Interestingly, there exist novel methods for so-called fairness repair that do not require the protected attribute \cite{zhou2023group,Abi2024} in the training data, but rather only the population-wide marginals of the protected attribute, which are widely available. While it may be impossible to quantify the amount of bias without access to the protected attribute, it may be possible to guarantee that any amount of bias has been compensated for using such methods.

Sample complexity also underlies the runtime complexity. While it is sometimes the case that the runtime is polynomial in the number of samples, if the sample complexity is high, the runtime may be high too.
Furthermore, there are distances such as Wasserstein-1 that are believed to be inapproximable \cite{lee2023computability}.

\section{Discussion}


Some of the major findings of the performed work, which should be taken into account in policy making for fairness AI are summarized next:
\begin{itemize}
    \item Discrimination by proxy variables, intersectional fairness and feedback loops comprise major issues when pursuing fairness in real world applications. Ongoing work is being performed in the algorithmic literature, although no one-size-fits-all solutions exist yet. Nevertheless, current legal frameworks need to be adapted and extended, so as to account for these particularities of the problem, that have emerged (partially) due to the uprise of research in algorithmic fairness.
    
    \item No one-size-fits-all fairness definitions or bias detection methods exist. Fairness is highly application-, scenario- and context-specific, since different real world applications of AI decision making systems and different social circumstances highly affect what is considered as ``fair''. Since the law cannot specialize on a case by case basis, this needs to be done by domain experts in collaboration with governmental and independent supervising and auditing authorities.
    
    \item Cross-sectorial collaboration is a necessity in practically every step of building both fair-by-design systems and methodologies, and AI fairness policies. There exists a large gap between law and ethics, and data and algorithms and only such collaboration can bridge this gap and produce meaningful policies and best practices.
    
    \item Some specific fairness definitions have been distinguished by a handful of prominent studies on the intersection of law and algorithms \cite{biaspreservationwachter2020bias, legalpersphauer2021legal}, e.g. Conditional Demographic Disparity, Equal Opportunity, Equalized Odds, Counterfactual Fairness, Calibration, can be considered suitable in different the application settings and contexts. Counterfactual Fairness is considered by part of the literature as a sufficiently expressive and adaptable definition that allows it to generalize in different cases and optimally represent substantive equality, in the spirit of the EU law.

\end{itemize}

The next steps of our work consist in performing a more thorough synthesis of best practices and recommendations identified in the relevant literature, at the intersection of algorithms and law and, based on this, propose a set of systematic guidelines for the design, deployment and assessment of fairness methods on AI systems, on real-world use cases.





\section*{Acknowledgment}
This work has been funded by the European Union’s Horizon Europe research and innovation
programme under Grant Agreement No. 101070568 (AutoFair).



\bibliographystyle{plain}
\bibliography{bibliography}

\end{document}